Organic Haze as a Biosignature in Anoxic Earth-like Atmospheres

Giada Arney[1,2]

NASA Goddard Space Flight Center

Mail Code 693

8800 Greenbelt Road

Greenbelt, MD 20771

Phone: 301-614-6245

giada.n.arney@nasa.gov

and

Shawn D. Domagal-Goldman[1,2], Victoria S. Meadows[2,3,4]

Affiliations:
[1]NASA Goddard Space Flight Center, 8800 Greenbelt Road, Greenbelt, MD 20771

[2]NASA Astrobiology Institute Virtual Planetary Laboratory, Box 351580, U.W. Seattle, WA 98195

[3]University of Washington Astronomy Department, Box 351580, U.W. Seattle, WA 98195

[4]University of Washington Astrobiology Program, Box 351580, U.W. Seattle, WA 98195

Abstract

Early Earth may have hosted a biologically-mediated global organic haze during the Archean eon (3.8-2.5 billion years ago). This haze would have significantly impacted multiple aspects of our planet, including its potential for habitability and its spectral appearance. Here, we model worlds with Archean-like levels of carbon dioxide orbiting the ancient sun and an M4V dwarf (GJ 876) and show that organic haze formation requires methane fluxes consistent with estimated Earth-like biological production rates. On planets with high fluxes of biogenic organic sulfur gases ($CS_2$, $OCS$, $CH_3SH$, and $CH_3SCH_3$), photochemistry involving these gases can drive haze formation at lower $CH_4/CO_2$ ratios than methane photochemistry alone. For a planet orbiting the sun, at 30x the modern organic sulfur gas flux, haze forms at a $CH_4/CO_2$ ratio 20% lower than at 1x the modern organic sulfur flux. For a planet orbiting the M4V star, the impact of organic sulfur gases is more pronounced: at 1x the modern Earth organic sulfur flux, a substantial haze forms at $CH_4/CO_2 \sim 0.2$, but at 30x the organic sulfur flux, the $CH_4/CO_2$ ratio needed to form haze decreases by a full order of magnitude. Detection of haze at an anomalously low $CH_4/CO_2$ ratio could suggest the influence of these biogenic sulfur gases, and therefore imply biological activity on an exoplanet. When these organic sulfur gases are not readily detectable in the spectrum of an Earth-like exoplanet, the thick organic haze they can help produce creates a very strong absorption feature at UV-blue wavelengths detectable in reflected light at a spectral resolution as low as 10. In direct imaging, constraining $CH_4$ and $CO_2$ concentrations will require higher spectral resolution, and R > 170 is needed to accurately resolve the structure of the $CO_2$ feature at 1.57 μm, likely the most accessible $CO_2$ feature on an Archean-like exoplanet.

1. Introduction

The Archean (3.8 – 2.5 billion years ago) eon may have experienced several intervals when a transient organic haze globally veiled our planet (e.g. Trainer et al. 2006; Zerkle et al. 2012; Izon et al. 2015; Hicks et al. 2016; Izon et al. 2017). This haze would have dramatically altered our planet's climate, spectral appearance, and photochemistry (Pavlov, Brown, et al. 2001; Pavlov, Kasting, et al. 2001; Domagal-Goldman et al. 2008; Haqq-Misra et al. 2008; Wolf & Toon 2010; Hasenkopf et al. 2011; Kurzweil et al. 2013; Claire et al. 2014; Arney et al. 2016; Arney et al. 2017). Organic haze formation is driven by methane photochemistry, and its optical thickness is controlled by the ratio of the amount of atmospheric methane ($CH_4$) relative to the amount of carbon dioxide ($CO_2$) (e.g. Trainer et al. 2006) because oxygen radicals produced by $CO_2$ photolysis can frustrate organic haze formation in Earth-like atmospheres (Arney et al. 2017).

On Archean Earth, there are numerous potential sources of methane, although biological processes likely dominated, as they do today (Kharecha et al. 2005). Methanogenesis, an anaerobic metabolism deeply rooted in the tree of life that likely evolved early in Earth's history (Woese & Fox 1977; Ueno et al. 2006), involves the uptake $CO_2$ and $H_2$ to form $CH_4$ and $H_2O$:

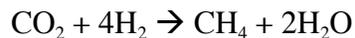
$$CO_2 + 4H_2 \rightarrow CH_4 + 2H_2O$$

Methanogenesis has only been observed in types of Archaea, and it occurs in a number of anoxic environments on modern Earth including animal guts and hydrothermal vents (Ver Eecke et al. 2012). In the latter environment, methanogens react $H_2$ contained in reduced vent fluids with $CO_2$ dissolved in the seawater. On Earth today, the flux of methane produced by biology is roughly $10^{11}$ molecules/cm$^2$/s (Pavlov, Brown, et al. 2001), and the Archean biotic flux has been estimated to range somewhere between 1/3 - 2.5 times this modern value (Kharecha et al. 2005).

The abiotic production rate of methane is not as well constrained. Serpentinization, the hydration of olivine and pyroxene, is the dominant abiotic source of methane on Earth today (Kelley et al. 2005; Guzmán-Marmolejo et al. 2013; Etiope & Sherwood Lollar 2013). The source of methane from serpentinizing systems is not completely clear. Serpentinization produces $H_2$, and it has been suggested that this $H_2$ can then react with $CO_2$ or CO to form $CH_4$ through Fischer-Tropsch type reactions (Bradley & Summons 2010). However, other explanations for methane produced in serpentinizing systems have been offered. For example, methanogens are known to live in these systems, and so some fraction of the methane produced from the vents is biological (Brazelton et al. 2006). However, based on the total density of cells expected in the vent systems, it is unlikely that methanogens are the dominant methane source in at least the Lost City hydrothermal system (Bradley & Summons 2010).  More recently, the isotopic composition of the hydrocarbons emanating from the Von Damm hydrothermal system were analyzed in relation to the isotopic composition of dissolved inorganic carbon (McDermott et al. 2015). Surprisingly, the isotopic composition of the vented $CH_4$ indicates that little-to-none of the $CH_4$ coming out the Van Damm vents is produced by reduction of dissolved inorganic carbon circulating through the rocks and participating in Fischer-Tropsch type chemistry. Possibly, the $CH_4$ produced in these systems is released from trapped magmatic fluid inclusions. Consistent with this finding, McCollom (2016) conducted laboratory experiments on the serpentinization of olivine with and without pyroxene using $^{13}$C-labeled $CO_2$ and found that production rates of $^{13}CH_4$ were inefficient. $^{13}$C-labeled $CH_4$ was produced in only one of their experiments: this experiment also contained a dissolved $H_2$-rich vapor phase that may help to promote Fischer-Tropsch type reactions. The kinetic barriers to methane formation in these reactions are surmountable in the presence of iron-nickel phase mineral catalysts such as awaurite, although the McDermott et al. (2015) study of the Von Damm vent implies that at least in some serpentinizing systems, these reactions are not important for the bulk methane production and another source dominates.

Kasting & Catling (2003) estimated the abiotic flux of methane as 1/300th the biotic rate. However, more recent measurements by Kelley et al. (2005) at the Lost City hydrothermal system indicate that abiotic methane production rates may actually be as high as 1/30th the biotic flux.

A different line of argument to estimate the abiotic methane production from serpentinization on Earth comes from Guzmán-Marmolejo et al. (2013). These authors calculate how much methane can be produced given current crustal spreading rates as a function of available FeO in the crust and by considering that $CO_2$ as the limiting reactant for methane production in serpentinziation – however, note that if the $CH_4$ is instead produced by a different mechanism, this assumption may longer apply. Based on the assumption of Fischer-Tropsch type synthesis, they estimate that the maximum amount of abiotic methane that can be produced from serpentinization is $6.8 \times 10^8$ molecules/cm$^2$/s for 1 Earth mass planet, which is only 1/160th the biotic flux (Guzmán-Marmolejo et al. (2013) also estimate the maximum $CH_4$ production rate as $1.3 \times 10^9$ molecules/cm$^2$/s for a five Earth mass planet). However, note that these results are contingent on the assumption of the modern-day Earth crustal spreading rate and the source of the methane.

Considering all of the above estimates, abiotic methane production estimates from serpentinization ranging between approximately 1/30 - 1/150th the present biotic flux appear reasonable for modern Earth. In the Archean, abiotic methane production rates are even less certain than they are today. If the early planet had faster seafloor spreading rates or a higher fraction of seafloor ultramafic rocks, enhanced abiotic methane compared to the modern planet would have been possible (Kasting 2005; Shaw 2008). Faster seafloor spreading rates may be more likely on a hotter young planet where convection may have proceeded more efficiently. How tectonics operated in the Archean remains uncertain, but there is geological and numerical modeling evidence supporting the existence of plate tectonics during this period (Kerrich & Polat 2006).

Although serpentinization is thought to be the dominant source of abiotic methane on Earth today, there are other abiotic methane sources (Etiope & Sherwood Lollar 2013). These sources include primordial delivery (via exogeneous sources); internal magmatic, postmagmatic and metamorphic processes; iron-carbonate decomposition; carbonate methanation; and aqueous $CO_2$ reduction. Global emissions of all non-anthropogenic biotic methane fluxes to the atmosphere have been well-studied and constrained at approximately 200 Mt/year (Pachauri et al. 2007). Total estimated methane emissions from geological sources amount to about 60 Mt/year (about $1\times10^{10}$ molecules/cm$^2$/s), although this flux is probably not purely abiotic (Etiope 2012): Etiope and Sherwood Lollar (2013) discuss how volcanic and non-volcanic geothermal systems can also release methane derived from thermal breakdown of sedimentary rock organic matter, which can be biotic (Etiope et al. 2007). The total global emission of truly abiotic methane is not well constrained. Emmanuel & Ague (2007) suggest it may be as low as ~2.3 Mt/year (about $4\times10^{8}$ molecules/cm$^2$/s), although Etiope and Sherwood Lollar (2012) point out that this estimate is hypothetical and not has not been constrained by direct measurements. Of course, we emphasize again that all of these estimates are for modern day Earth; past Earth, and other planets, will naturally have different methane flux rates.

Given all of this, finding an organic haze in the atmosphere of a planet with Archean-like $CO_2$ levels could be indicative of highly interesting processes that imply ongoing geological activity and/or biological methane production with high methane source fluxes to drive haze-production in a $CO_2$-rich atmosphere. Such a planet should be a target of closer follow-up studies aimed at discriminating between geological and biological $CH_4$ sources and to search for other signs of habitability and life (Section 4.3). We emphasize that while detection of $CH_4$ and organic haze – which can have a significantly stronger spectral signature than $CH_4$ – in an Earth-like atmosphere would be a tantalizing hint of the presence of methane-producing life on an exoplanet (since biology produces the bulk of methane on modern Earth), it would not be enough to conclude biological activity given the existence of abiotic $CH_4$ sources. This is similar to – but perhaps more obvious than – the case for oxygen as a biosignature given that several pathways for abiotic oxygen production have recently come to light (Domagal-Goldman et al. 2014; Luger & Barnes 2015; Harman et al. 2015; Schwieterman et al. 2016). The spectral signatures of a possibly biologically-produced spectral features need to be placed in the context of a broader understanding of that planet's atmospheric chemistry and potential for habitability. In this manuscript, we pursue one type of contextual information that would help to discriminate between abiotic and biotic hazes on Earth-like worlds.

It has been pointed out that biogenic organic sulfur gases ($S_{org}$) can contribute to the atmospheric hydrocarbon budget through photochemical processes that liberate organic species from the $S_{org}$ molecules (Domagal-Goldman et al. 2011). These $S_{org}$ gases include carbon disulfide ($CS_2$), carbonyl sulfide (OCS), methanethiol ($CH_3SH$, also called methyl mercaptan), and dimethyl sulfide ($CH_3SCH_3$ or DMS). Volcanic processes can also produce $CS_2$ and OCS, but at lower fluxes than Earth's biology (Lomans et al. 2002). The potential for $S_{org}$ gases to act as biosignatures has been considered by previous studies (Pilcher 2003; Vance et al. 2011), and Domagal-Goldman et al. (2011) showed that although $S_{org}$ gases may be difficult to directly detect in a planet's spectrum, their photochemical byproducts can produce spectral signatures that can indirectly imply a flux of these gases. In particular, Domagal-Goldman et al. (2011) found the photolysis of $S_{org}$ gases can release methyl radicals ($CH_3$) that contribute to ethane ($C_2H_6$) production in excess of the amount predicted from methane photochemistry alone. This effect is especially pronounced around M dwarfs whose UV spectral output allows for longer atmospheric lifetimes of $C_2H_6$ than solar-type stars. Unfortunately, $C_2H_6$ absorbs most strongly in the mid-infrared at 12 μm, making its detection potentially difficult. Domagal-Goldman et al. (2011) did not consider $S_{org}$-rich atmospheres with enough methane to lead to haze formation, nor did they simulate the spectral region in which hazes have detectable features. But the same principles that caused higher $C_2H_6$ in their simulations could also cause a greater haze concentration in the presence of $S_{org}$. Here, we will test whether the hydrocarbons contributed to the atmosphere by $S_{org}$ photochemistry can induce haze formation at lower $CH_4/CO_2$ ratios than would be expected if haze formation was driven by methane production alone, thereby providing a spectral clue that biological activities may be influencing haze formation on a planet. Organic haze is a particularly useful potential biosignature because it produces a very strong broadband absorption feature at ultraviolet (UV) and visible wavelengths (this is the reason why Titan is orange), and it also produces absorption features in the NIR. These features may be accessible with observatories becoming available in the coming decades, including the James Webb Space Telescope (JWST) and possible future large direct-imaging telescopes such as LUVOIR and HabEx (Postman et al. 2010; Bolcar et al. 2015; Dalcanton et al. 2015; Mennesson et al. 2016) as we have studied previously (Arney et al. 2017).

2. Methods

To simulate Archean-analog planets, we use a coupled 1D photochemical-climate model called Atmos. The Atmos model is described in detail in Arney et al. (2016), and limitations of the Atmos haze formation scheme are discussed in Arney et al. (2016) and Arney et al. (2017). In brief, this model assumes a pathway proposed for Titan's hazes (Allen et al. 1980; Yung et al. 1984) where haze formation occurs via polymerization of acetylene ($C_2H_2$). In this scheme, haze particles are formed via $C_2H + C_2H_2 \rightarrow C_4H_2$ and $C_2H + CH_2CCH_2 \rightarrow C_5H_4 + H$. However, *Cassini* measurements of Titan show that haze formation is more complex, and includes ion chemistry and the formation of nitriles (Waite et al. 2007; Vuitton et al. 2009; López-Puertas et al. 2013). Additionally, while Titan's atmosphere is extremely reducing, Archean Earth's atmosphere would have been less so, and laboratory studies have shown that oxygen atoms can be incorporated into haze molecules (Trainer et al. 2006; DeWitt et al. 2009; Hörst & Tolbert 2014; Hicks et al. 2016). Lack of these processes in Atmos may cause the model to underpredict the haze formation rate; on the other hand, in a real atmosphere, $C_4H_2$ would be able to revert back to $C_2H_2$. Our model does not include this, which could lead to haze over-prediction. Ongoing improvements to Atmos will include a more complete haze formation scheme.

The climate portion of Atmos was originally developed by Kasting & Ackerman (1986), although it has been significantly modernized since then, and it was most recently updated and described in a re-calculation of habitable zone boundaries around main sequence stars (Kopparapu et al. 2013) and in a study of the impact of organic haze in the Archean (Arney et al. 2016). The photochemical portion of Atmos is based on a code developed by Kasting et al. (1979), and it was significantly modernized by Zahnle et al. (2006). This model, supported by the Virtual Planetary Laboratory in NASA's Astrobiology Institute, is now publicly available at https://github.com/VirtualPlanetaryLaboratory/atmos.

The climate model is considered converged when the change in flux out the top of the atmosphere and change in surface temperature are sufficiently small (typically on the order of $1\times10^{-5}$), and when the energy from the star into the atmosphere balances the energy radiated out of the atmosphere. The photochemical model uses a first order reverse Euler solver to solve continuity and flux equations for each species at all altitudes. In the timestepping loop, the model tracks how much the gas concentrations change in each step, and the species with the largest relative error in its change in concentration (called $E_{max}$) is used to set the size of the next timestep. When $E_{max}$ is small, the next timestep will be larger; if $E_{max}$ is too large, the model will decrease the timestep size. The model checks the timestep length to determine convergence, and when the timestep size exceeds $1\times10^{17}$ seconds, the model considers itself "converged" and stops.

Both the climate and photochemical models have been modified to simulate haze particles as fractal in shape rather than as spherical (Mie) particles (Wolf & Toon 2010; Zerkle et al. 2012; Arney et al. 2016) using the fractal mean field approximation (Botet et al. 1997). Studies of Titan's atmosphere indicate that fractal particles are more realistic for organic hazes (Rannou et al. 1997), and early Earth analog fractal organic hazes have been simulated in the laboratory (Trainer et al. 2006). Zerkle et al. (2012) and Arney et al. (2016) describe our model's haze particle treatment in detail. Fractal particles are composed of multiple smaller spherical particles called monomers clumped together into complex branching forms, and their scattering and absorption physics differ from spherical particles. In general, compared to equal mass spherical particles, fractal particles produce more extinction at shorter wavelengths, and less extinction at longer wavelengths. The consequences of this behavior have been previously considered in the context of UV-shielding and climate cooling effects for an Archean haze (Wolf & Toon 2010; Zerkle et al. 2012; Arney et al. 2016). Generally, haze particles initially form at altitudes of 80-90 km; this altitude is a result of the model's photochemistry and is not prescribed.

The methyl radicals produced by $S_{org}$ gases that participate in haze formation photochemistry can be generated by reactions such as:

$$CH_3SH + O \rightarrow CH_3 + HSO \quad (1)$$

$$CH_3SH + h\nu \rightarrow CH_3 + HS \quad (2)$$

The full chemical network that $S_{org}$ gases participate in is discussed in detail in Domagal-Goldman et al. (2011). Once $CH_3$ is produced, it directly contributes to haze formation via the process outlined in Arney et al. (2017): $CH_3$ produces ethane most efficiently through $CH_3 + CH_3CO \rightarrow C_2H_6 + CO$. Ethane can then be photolyzed to produce $C_2H_4$ or $C_2H_2$, or it can react with OH to produce $C_2H_5$, all of which step towards haze formation.

Spectra are generated using the Spectral Mapping Atmospheric Radiative Transfer model (SMART) (Meadows & Crisp 1996; Crisp 1997) using outputs from the Atmos model. SMART is a 1D, line-by-line, fully multiple scattering radiative transfer model. Haze is included in SMART from Atmos via a particle binning scheme described in Arney et al. (2016). The newest version of SMART can also calculate transit transmission spectra in the same model run that calculates reflected light spectra. The model's transit calculations include the path length and refraction effects inherent in transit transmission spectra (Misra, Meadows & Crisp 2014; Misra, Meadows, Claire, et al. 2014).

To simulate observations with possible large future space-based telescopes, we use the coronagraph noise model described in Robinson et al. (2016). The same nominal parameters are assumed as those discussed in Robinson et al., except we assume a telescope operating temperature of 270 K and a constant quantum efficiency as a function of wavelength (0.9). Noise sources include: dark noise, read noise, zodiacal and exo-zodical light, stellar light leackage, telescope thermal radiation. A publically accessible online version of this simulator is available at https://asd.gsfc.nasa.gov/luvoir/tools/.

2.1 Model Inputs

We compare haze production under the influence of $S_{org}$ gases for planets orbiting the sun 2.7 billion years ago (Ga) during the Archean eon and the M4V dwarf GJ 876 with the same total insolation as the 2.7 Ga sun (0.8 x 1360 W/m$^2$). GJ 876 is a known multi-planet host (Von Braun et al. 2014). The spectrum we use for it is described in Domagal-Goldman et al. (2014) based on the spectrum reported in France et al. (2012). We chose this star over higher activity M dwarfs (e.g. AD Leo) because Domagal-Goldman et al. (2011) showed that $S_{org}$ gases have a greater impact on hydrocarbon photochemistry for lower-activity M dwarfs. Stars with lower UV outputs generate fewer photochemical oxygen species, which are major sinks of both hydrocarbons and organic sulfur gases (Domagal-Goldman et al. 2011, Arney et al. 2017). For the sun, we corrected its spectrum for higher levels of expected activity when it was younger using the wavelength-dependent solar evolution correction from Claire et al. (2012), which estimates the solar flux at different epochs by combining data from the sun and solar analogs to determine appropriate wavelength-dependent stellar flux corrections. The stellar spectra used in this study are shown for the UV, visible, and near-infrared (NIR) in Figure 1. The UV spectra shown in Figure 1 show the actual resolution of the wavelength grid used by the photochemical model. The model's "Lyman alpha" bin encompasses flux from wavelengths spanning 8 angstroms wide on either side of Lyman alpha (121.6 nm).

Our chemical reaction network is based on the one used by Arney et al. (2016) and Arney et al. (2017), although these earlier studies did not include $S_{org}$ gases. The supplementary online information of Arney et al. (2016) provides a complete list of chemical reactions and species boundary conditions for this nominal Archean model. To include $S_{org}$, we updated our templates based on the reaction list and boundary conditions discussed in Domagal-Goldman et al. (2011). However, we removed the $NH_3$-related gases and reactions from the templates shown in Domagal-Goldman et al. (2011) because we have not yet incorporated $NH_3$ into our climate model, and it is a potentially significant greenhouse gas. Inclusion of $NH_3$ remains an important area of future work because it may play a role in warming our early planet under the fainter young sun, especially under a haze that could protect it from UV-photolysis (Sagan & Chyba 1997; Pavlov et al. 2001; Wolf & Toon 2010). Our atmospheres with 1x $S_{org}$ fluxes use the same $S_{org}$ surface boundary fluxes presented in Table 2 of Domagal-Goldman et al. (2011). In units of molecules/cm$^2$/s, these fluxes are: $1.4 \times 10^7$ for $CS_2$ and OCS, $8.3 \times 10^8$ for $CH_3SH$, $4.2 \times 10^9$ for $CH_3SCH_3$, and 0 for $CH_3S_2CH_3$. This last species, $CH_3S_2CH_3$ or dimethyl disulfude (DMDS), is not produced by biology but results from $S_{org}$ photochemistry. Other non-biological $S_{org}$ gases relevant to $S_{org}$-photochemistry included in our photochemical scheme are CS ($1.7 \times 10^7$ molecules/cm$^2$/s produced at the surface), and $CH_3S$ (0 molecules/cm$^2$/s produced at the surface, but photochemistry can produce this gas in the atmosphere).

Unlike our previous studies where we set the $CH_4$ surface mixing ratio to explore haze formation under different $CH_4$ concentrations (Arney et al. 2016, Arney et al. 2017), here we vary the $CH_4$ surface flux and allow the photochemical model to calculate the self-consistent atmospheric mixing ratio from the selected fluxes to explore haze formation under different $CH_4$ production rates. The $CO_2$ atmospheric fractions ($fCO_2$) simulated here range from $1 \times 10^{-5}$ to $1 \times 10^{-1}$; the lower limit on $CO_2$ abundance was chosen to roughly represent the limit for C4 photosynthesis (Kestler et al. 1975; Tolbert et al. 1995), which is determined by the ability of plant stomata to maintain a diffusive gradient of $CO_2$ concentration from the atmosphere into the cellular structure. Estimates of Archean $CO_2$ have ranged from values close to those of modern Earth to orders of magnitude higher (e.g. Rosing et al. 2010; Dauphas & Kasting 2011; Driese et al. 2011; Kanzaki & Murakami 2015); in our previous work (Arney et al. 2016, 2017), we adopted the values from Driese et al. (2011) for our nominal estimates of pCO$_2$ at 2.7 Ga (pCO$_2$ ~ $1 \times 10^{-3} - 1 \times 10^{-2}$ bar). Here, we choose an upper limit on $CO_2$ that is an order of magnitude larger than the Driese et al. (2011) range. Note that organic haze formation in a much more oxidizing atmosphere (such as Mars-like with 95% $CO_2$ or modern day Earth-like with its 21% $O_2$) is not tenable at any plausible hydrocarbon production rates. For significantly more reducing atmospheres than those shown here (i.e. Titan-like), haze formation can be possible at very low methane source fluxes compared to the ones we simulate, which would make an argument for biological involvement in methane production difficult. In this study, methane surface fluxes were chosen to range between $6.8 \times 10^8$ to $1 \times 10^{12}$ molecules/cm$^2$/s. The lower limit on methane production is taken from the theoretical study of abiotic, serpentinization-driven methane production by Guzmán-Marmolejo et al. (2013) for Earth-like worlds. Life on Earth produces $CH_4$ at a rate of $\sim 1 \times 10^{11}$ molecules/cm$^2$/s, and the upper limit for methane flux we consider is an order of magnitude larger than this amount.

For haze optical properties, we use the optical constants of Khare et al. (1984) subject to the caveats outlined in Arney et al. (2016), where we discuss how these optical constants were derived for Titan-analog (not Archean-analog) hazes. However, Archean-simulant haze optical constants have only been measured at one wavelength (532 nm) by a previous study (Hasenkopf et al. 2010), and the Khare et al. (1984) haze measurements agree reasonably well with the Hasenkopf et al. (2010) measurement. We are currently involved with laboratory work to simulate and measure new optical constants for Archean-analog organic hazes from the UV to the NIR, but haze production rates are slow in $CO_2$-rich conditions and the analyses will not be ready in time for inclusion in this manuscript. However, we will use these updated optical constants in the future and make them publically available once we do.

We assume a total surface pressure of 1 bar for all simulations. The background atmosphere is composed of $N_2$. When we refer to $CH_4/CO_2$ ratios, note that we are referring to the value at the surface since $CH_4$ does not follow an isoprofile in these atmospheres. $CO_2$, meanwhile, is assumed to be well-mixed. We set molecular oxygen ($O_2$) at a mixing ratio of $1 \times 10^{-8}$, corresponding to a time after the origin of oxygenic photosynthesis but before substantial oxygen accumulation in the atmosphere (Kharecha et al. 2005; Claire et al. 2014). Note that haze can form at higher oxygen concentrations than considered here, and possibly even at oxygen concentrations corresponding to the low Proterozoic $O_2$ levels suggested by Planavsky et al (2014) of 0.1% the present atmospheric level as shown by Kurzweil et al. (2013) and Izon et al. (2017). However, haze can only form in such atmospheres at $CH_4$ fluxes higher than those we consider here.

To generate spectra, we use the HITRAN 2012 linelists (Rothman et al. 2013) and set a solar zenith angle of 60°, which approximates the flux observed at quadrature. As in Atmos, optical constants for the haze particles in SMART are derived from the Khare et al. (1984) optical constants using the fractal mean field approximation.

3. Results

In this section, we consider haze as a biosignature in the context of atmospheres with and without $S_{org}$ gases.

As a baseline case, we first consider haze formation in the absence of biogenic sulfur gases to explore how varying $CO_2$ and $CH_4$ levels affects haze production. Figure 2 shows the optical thickness of an organic haze in the atmosphere an Archean planet with no $S_{org}$ orbiting the ancient (2.7 Ga) sun, and Figure 3 shows the same for an Archean-analog planet orbiting GJ 876. In these contour plots, optical depth of unity at 190 nm (chosen to represent hazes that would significantly impact photochemistry and the planet's spectrum) is marked with the solid black lines overlying the colored contours. Figures 2 and 3 show that lower $CH_4$ fluxes are needed to form optically thick hazes for the simulated planets around GJ 876 compared to the planets simulated around the ancient sun. This is consistent with our findings in Arney et al. (2017), which showed that lower $CH_4/CO_2$ ratios are required to form hazes around Archean-analog worlds orbiting GJ 876, because this star's lower UV output generates fewer haze-destroying oxygen radicals from processes like $CO_2$ photolysis. At the Driese et al (2011) $CO_2$ levels, haze production requires methane fluxes broadly consistent with estimated Archean biological methane production rates ($\sim 0.3 - 2.5 \times 1 \times 10^{11}$ molecules/cm$^2$/s) according to Kharecha et al. (2005). This is the case for both simulations of a planet orbiting the Archean sun and simulations of a planet orbiting GJ 876.

In Figures 4-7, we include $S_{org}$ fluxes to show how this additional source of hydrocarbons affects haze formation. Figures 4 and 6 are analogous to Figures 2 and 3 for the sun and GJ 876, respectively. In the gray region of Figures 6 and 7, simulations are not converged; in this part of parameter space, very thick hazes generated extreme stratospheric heating (~400 K), causing model instabilities. In Figures 5 and 7, we show the same data but here, the horizontal axis shows the $CH_4/CO_2$ ratio instead of the $CH_4$ flux for two reasons. First, this emphasizes how the presence of significant $S_{org}$ fluxes can impact the $CH_4/CO_2$ ratio required to form haze. Second, these plots show the actual spectral observables since it is the $CH_4$ mixing ratio – not its flux – that can be observed from a spectrum (the latter could only be inferred through modeling once the concentrations of atmosphere gases – and the UV spectrum of the star – were measured). In all four figures of simulations including $S_{org}$ fluxes (Figures 4-7), the top panel corresponds to the optical depth contours for 1x the modern $S_{org}$ fluxes as defined in Domagal-Goldman et al. (2011), and the bottom panel corresponds to 30x the modern $S_{org}$ fluxes, chosen for consistency with the upper limit for the $S_{org}$ flux value in Domagal-Goldman et al. (2011). In both panels, the solid red line denotes where the haze's 190 nm optical depth equals unity for 1x $S_{org}$, and the solid black line denotes where the haze optical depth is unity for 30x $S_{org}$. These lines are plotted together in both panels so that they can be easily compared.

Figures 5 and 7 show that as $fCO_2$ decreases, the $CH_4/CO_2$ ratio required to form a thick haze increases. This result is counter-intuitive and arises because the total carbon budget of the atmosphere decreases as $CO_2$ is removed; in other words, any given atmosphere has less $CH_4$ available to form haze at a fixed $CH_4/CO_2$ ratio at lower $CO_2$ levels. More significantly, when fluxes of $S_{org}$ increase, the $CH_4/CO_2$ ratio necessary to form a haze decreases at all $CO_2$ levels for planets orbiting both stars. This occurs because photochemistry readily forms methyl groups from organic sulfur gases, and formation of $CH_3$ is a step in the haze formation process. By increasing the efficiency of these processes, the $CH_4/CO_2$ ratio required to form haze at a given $CO_2$ concentration is lowered.

For a solar-type star with 1x $S_{org}$, Figure 4 shows the $CH_4$ fluxes required to form haze are similar to the simulations with no $S_{org}$. However, in the presence of 30x $S_{org}$, the $CH_4$ fluxes required to initiate haze formation decrease by about 50% for the highest $CO_2$ level considered ($fCO_2 = 10^{-1}$) and by almost an order of magnitude for $fCO_2 = 10^{-3}$. Figure 5 shows that at $fCO_2 = 10^{-2}$, the $CH_4/CO_2$ ratio needed to form a haze for the planet around the solar-type star with $\tau(190\ nm) = 1$ at 190 nm is about 0.2 for 1 x $S_{org}$ case. This is roughly the same as the $CH_4/CO_2$ ratio required to form a substantial haze in the absence of $S_{org}$ (Arney et al. 2016). For the 30x$S_{org}$ case, the $CH_4/CO_2$ ratio required to form substantial haze for the same planet with $fCO_2 = 10^{-2}$ is about 0.16, a 20% decrease.

Around GJ 876, $S_{org}$ has a larger impact on haze formation with markedly less methane required to form an optically thick haze at high $S_{org}$ fluxes. In Figure 6, at $fCO_2 = 10^{-2}$, the haze becomes optically thick at over an order of magnitude smaller $CH_4$ fluxes in the presence of 30 x $S_{org}$ compared to 1 x $S_{org}$. Figure 7 shows that at 1 x $S_{org}$, the haze becomes optically thick at slightly less than $CH_4/CO_2 = 0.2$, but for 30 x $S_{org}$, the haze becomes optically thick at $CH_4/CO_2 \sim 0.02$, a full order of magnitude lower. As discussed above, haze forms more readily in the atmosphere of the GJ 876 planet because GJ 876's spectrum results in the production of a smaller quantity of haze-destroying oxygen radicals from $CO_2$ photolysis compared similar planets orbiting the sun (Arney et al. 2017).

Figures 5 and 7 show that for planets around both stars, as $fCO_2$ decreases (and, therefore, as the absolute amount of $CH_4$ in the atmosphere at a given $CH_4/CO_2$ ratio also decreases), the difference between the $CH_4/CO_2$ ratios required to form a thick haze with $1xS_{org}$ versus $30xS_{org}$ increases because $S_{org}$ becomes a larger proportional contributor to the atmosphere's hydrocarbon budget.

Interestingly, $S_{org}$ gases themselves provide enough hydrocarbons to not only affect haze-formation but also affect the absolute methane abundance. For example, for a planet around the sun with a methane flux of $1x10^{11}$ molecules/cm$^2$/s and $fCO_2 = 1x10^{-2}$, we find that an atmosphere without $S_{org}$ gases generates a surface methane mixing ratio of $2.2x10^{-4}$, while an atmosphere with $30xS_{org}$ gases generates a surface methane mixing ratio of $6.8x10^{-4}$. Around GJ 876, the same planets have surface methane mixing ratios of $6.7x10^{-4}$ and $7.3x10^{-3}$ for no $S_{org}$ and $30xS_{org}$, respectively. The largest photochemical sources of $CH_4$ in the $30xS_{org}$ atmospheres are $CH_3 + HCO \rightarrow CH_4 + CO$ and $CH_3 + H \rightarrow CH_4$. These reactions occur one to two orders of magnitude faster in the $30xS_{org}$ atmospheres due to the production of $S_{org}$-derived methyl radicals. Although the high $S_{org}$ atmospheres have higher methane levels, which itself allows haze to form more readily, Figures 5 and 7 show clearly that the hazes in the high $S_{org}$ atmospheres still form at lower methane mixing ratios than they would without $S_{org}$ gases because the $S_{org}$ gases also contribute other gases relevant to haze formation (e.g. $CH_3$ and $C_2H_6$). The methane derived from $S_{org}$ gases also imposes a lower limit on the amount of methane in these high $S_{org}$ atmospheres. For instance, the lowest $CH_4/CO_2$ ratios generated in our $30xS_{org}$ simulations (bottom panels of Figures 5 and 7) are higher than the lowest $CH_4/CO_2$ ratios for $1xS_{org}$ (top panels of Figures 5 and 7).

4. Discussion

Haze formation in atmospheres with Archean-like levels of $CO_2$ can indicate methane production rates consistent with known and theoretical Earth-like biogenic methane production rates, and these fluxes are higher than known and theoretical rates of abiotic methane production on Modern Earth. However, more efficient abiotic methane production rates on early Earth and exoplanets cannot be ruled out. Thus, while haze would not be definitive proof of life on these planets, it would indicate that they are consistent with the behavior of Earth's methanogenic biosphere and therefore are highly interesting targets for follow-on studies. We have shown here that detecting an organic haze in the presence of a relatively low $CH_4/CO_2$ ratio could further imply the influence of biogenic sulfur gases aiding haze formation. This is similar to the suggestion of ethane as a spectral biosignature in Domagal-Goldman et al. (2011), as photolysis of methyl-bearing $S_{org}$ gases enhances formation of ethane in atmospheres with large $S_{org}$ fluxes even if the $S_{org}$ gases themselves are difficult to detect directly. The impact of $S_{org}$ on haze formation is more pronounced around M dwarf stars like GJ 876 because the lower overall UV flux around this star leads to smaller sinks of hydrocarbon gases (Domagal-Goldman et al. 2011, Arney et al. 2017).

In all situations – including planets without $S_{org}$ gases – it will be important to know the redox state and temperature regime of an atmosphere to argue for the plausible biogenecity of an organic haze. Titan shows us that abiotic hazes can form in highly reducing cold atmospheres with long residence lifetimes for methane. Estimating the temperature of a planet will require measurements of the planet's semi-major axis and greenhouse gas budget. However, the temperature difference between Titan and early Earth is large – about 200 K – and so only broad temperature constraints would be needed to separate Titan-like planets from Earth-like planets. To understand the atmospheric redox state of a given world, measurements of $CO_2$ and $CH_4$ will be required to constrain the $CH_4/CO_2$ ratio. Additional discussion on the interpretation of a haze spectral feature in the context of the rest of the planetary environment can be found in Section 4.3.

4.1 Detectability Considerations

Transit observations with JWST (Beichman et al. 2014) could provide access to NIR wavelengths on hazy exo-Earths and were discussed in detail Arney et al. (2017). Planets orbiting M dwarfs are more amenable to observations with JWST compared to planets around solar-type stars due to the larger planet-to-star size ratio and more frequent transits for planets in the habitable zone. In Figure 8 we show the transit transmission spectra of hazy Archean Earth with 1x and 30x $S_{org}$ around the sun and GJ 876 to illustrate which spectral features may be detectable for these worlds. The planets shown around the sunlike star were simulated with a methane flux of $6.9 \times 10^{10}$ molecules/cm$^2$/s and a $CO_2$ mixing ratio of $1 \times 10^{-3}$, corresponding to surface methane mixing ratios of $7.9 \times 10^{-5}$ and $3.2 \times 10^{-4}$ for 1x and 30x $S_{org}$, respectively. The planets around the GJ 876-like star were simulated with a methane flux of $3.1 \times 10^{10}$ molecules/cm$^2$/s and a $CO_2$ mixing ratio of $1\times10^{-2}$, with resultant surface methane mixing ratios of $1.4\times10^{-4}$ and $1.2\times10^{-3}$. These atmospheres were chosen to represent cases where haze is not significantly present at 1x $S_{org}$ but is present at 30x $S_{org}$. As expected, Figure 8 shows that the spectral consequences of high $S_{org}$ fluxes are more significant for the planet orbiting GJ 876.

Even at 30x $S_{org}$, organic sulfur gases are not apparent in the transit transmission spectrum. Domagal-Goldman et al. (2011) showed that these $S_{org}$ gases are concentrated in the lower atmosphere since they are readily photolyzed at higher layers, and so they are spectrally inaccessible at the altitudes probed by transits for a hazy planet (20-80 km for the planets in Figure 8). Haze, $CH_4$, $CO_2$, and $C_2H_6$ are all potentially detectable, however. Organic haze could be discerned through the presence of a haze-induced NIR spectral slope, and haze produces an absorption feature near 6 μm, and a much weaker one near 3 μm (not labeled). $CH_4$ absorbs near 1.1, 1.4, 1.7, 2.3, 3.3, and 7.5 μm. $CO_2$ also absorbs at wavelengths accessible to JWST at 1.57, 2, 2.7, and 4.3 μm. $C_2H_6$ produces features near 6.5 and 12 μm. The features at wavelengths longward of about 8 μm can probably be considered inaccessible to JWST due to the dim stellar blackbody at these wavelengths. See Arney et al. (2017) for our discussion of how observable these features are with JWST for a planet orbiting a star like GJ 876.

Potential direct imaging telescopes under consideration such as the Large UV Optical Infrared telescope (LUVOIR) and the Habitable Exoplanet Imaging Mission (HabEx) may provide direct observations of Earth-like exoplanets in the 2030s and beyond (Postman et al. 2010; Dalcanton et al. 2015; Bolcar et al. 2015; Stapelfeldt et al. 2015;

Mennesson et al. 2016). Figure 9 shows the reflected light spectra of the same planets shown in Figure 8. Haze produces a broad, deep spectral feature at UV-blue wavelengths; for this combination of $CH_4$-$CO_2$-$S_{org}$, the haze feature is weaker for the planet around the sunlike star compared to the planet around the GJ 876-like star, but it still significantly alters the shape of the spectrum at wavelengths < 0.5 μm. This very strong feature is the reason why we argue here that haze is likely a much more detectable indicator of atmospheres with $S_{org}$ compared to the $C_2H_6$ discussed in Domagal-Goldman et al. (2011) that absorbs in the mid-IR. Of course, for both transit transmission and direct imaging observations, how well we will be able to determine whether a haze is present at an anomalously low $CH_4/CO_2$ ratio depends on how well we will be able to retrieve the $CH_4$ and $CO_2$ gas concentrations.

To examine the spectral resolutions (R = $\lambda/\Delta\lambda$) required to observe features in hazy reflected light spectra, Figure 10 shows the reflectance spectrum of a representative hazy Archean Earth at several spectral resolutions for $fCO_2 = 10^{-2}$ and $CH_4/CO_2 = 0.2$. The Figure 10 spectrum includes water clouds added to our 1D radiative transfer model using a weighted averaging technique where 50% of the planet is considered haze-covered and cloud-free, 25% is covered by haze and stratocumulus clouds, and 25% by cirrus clouds and haze (Robinson et al. 2011). Haze produces a broad, strong absorption feature at short wavelengths that can be easily resolved at spectral resolutions as low as 10. Methane and $CO_2$ are observable in the NIR, although $CO_2$ is more challenging to detect. Carbon dioxide has features near 1.57 and 2 μm for the wavelengths shown here, but telescope thermal emission could swamp the 2 μm feature for a non-cryogenically cooled mirror. Therefore, the most detectable $CO_2$ feature in reflected light occurs at 1.57 μm for Archean-like $CO_2$ levels. Resolving the narrow multi-peaked structures of the 1.57 μm $CO_2$ feature will require spectral resolution R > 100, and R > 170 will be required to correctly resolve the depths of the narrow bands of this feature. As in the transit transmission spectrum, $S_{org}$ features are not detectable in the reflected light spectrum.

Although there are numerous interesting spectral features in the NIR, they may not be detectable in direct imaging observations even if telescopic thermal emission is negligible because inner working angle (IWA) constraints alone can limit access to longer wavelengths. The IWA represents the smallest planet-star angular separation that can be resolved with sufficient signal-to-noise to detect the planet. For a coronagraph, the IWA scales with C×λ/D where C is a small valued-constant of order unity, and D is the telescope diameter. Several mirror diameters are being considered for the designs of the HabEx and LUVOIR mission concepts. Currently, these designs include a 4m monolith (HabEx), a JWST-sized 6.5 m mirror (HabEx), a 9.2 m mirror (LUVOIR), and a 15.1 m mirror (LUVOIR). Note, however, that light reflected off the jagged hexagonal segments on the outside of a segmented mirror poses a challenge for current coronagraph designs. Therefore, to be conservative, we will consider only light reflected from the largest inscribed circle that fits within each segmented aperture, resulting in a smaller effective aperture visible to the coronagraph. For the segmented telescope sizes listed previously, these correspond to inscribed diameters of 5.5 m, 7.6 m, and 12.7 m, respectively, for the current designs of these telescopes (M. Bolcar, personal communication).

To receive an Archean-like level of instellation (i.e. stellar irradiation), a planet orbiting a star like GJ 876 would be at a planet-star separation distance of 0.12 AU. Assuming an optimistic IWA of λ/D and a 12.7 m aperture, a GJ 876-like planet-star system could be located no farther away than 4.5 parsecs (pc) for the IWA to allow characterization of the spectrum out to 1.57 μm to see the $CO_2$ band using a coronagraph. Smaller telescopes fare worse, as does assuming more conservative IWAs such as IWA = 2 λ/D or IWA = 3 λ/D. Table 1 shows the longest wavelength observable before the IWA cuts off the spectrum for planet-star systems located at a fixed distance of 4.5 pc assuming a solar-type star and a GJ 876-type star for the four different telescope sizes being studied by HabEx and LUVOIR (with the IWAs for the segmented telescopes use their inscribed circle diameters). At 4.5 pc, it is difficult to characterize the planet orbiting GJ 876 in the NIR unless the telescope is the largest we simulate and/or the IWA is the most optimistic we consider. The influence of $S_{org}$ on haze formation is most pronounced around M dwarfs like GJ 876, but as we have seen, IWA constraints make Earth-like planets around such stars more challenging to characterize since they orbit so close to their host stars. Note, however, that here we are assuming that starlight suppression is achieved with a coronagraph, but starshades may be able to provide smaller IWAs than those discussed here, depending on starshade size and starshade-telescope separation distance (Turnbull et al. 2012).

Figure 11 shows the integration times required to obtain a signal-to-noise ratio (SNR) of 20 as a function of wavelength for planets with the same hazy Archean atmospheric parameters used to generate Figure 10 using the Robinson et al. (2016) coronagraph noise model. We assume R = 170 in the visible and NIR to fully resolve the structure of the narrow 1.57 µm $CO_2$ band peaks and troughs, and R = 20 in the UV (for $\lambda < 0.4$ µm). Vertical lines indicate IWA cutoffs as described in the figure caption. We again assume apertures of 4 m, 5.5 m, 7.6 m, and 12.7 m, and the planet-star system is assumed to be 4.5 pc to allow detection of the 1.57 µm $CO_2$ band for a planet orbiting a star like GJ 876 with IWA = $\lambda/D$ for the largest telescope. We chose to simulate SNR = 20 because studies have shown that continuum SNR measurements of about 20 may be required to constrain gas abundances in planetary atmospheres (Lupu et al. 2016; Nayak et al. 2017, M. Marley personal communication). Note in Figure 11 that absorption bands, where the planet is darker, require longer integration times to reach SNR = 20, but the gas retrieval demands discussed above apply to the continuum regions, where the planet is brighter. Because M dwarfs are dimmer than sunlike stars in the visible, longer integration times are required to characterize the planet around GJ 876 at wavelengths < 1 µm compared to the planet orbiting the sun. The situation is reversed for wavelengths > 1 µm. The dramatic increase in integration times near 1.6 µm is caused by the telescope's thermal emission (the mirror is 270 K), rendering these longer wavelengths effectively unobservable for a non-cryogenic telescope even if they were accessible within the IWA wavelength cutoff.

Integration times at all wavelengths increase by at least an order of magnitude for the 4 m telescope compared to the 12.7 m telescope. For the 12.7 m telescope, sensing continuum regions to a level of SNR = 20 would generally require 10s-100s of hours of observing time for the resolution shown here. For the 4 m telescope, measuring the continuum at SNR = 20 demands 100s-1000s of hours of integration.

4.2 Other Hydrocarbon-Containing Gases

Although we included just $S_{org}$ gases in our photochemical model, there are numerous other methyl-containing gases produced by terrestrial organisms and other processes (Seager et al. 2012) that may contribute to photochemical organic haze production if present in a planet's atmosphere. Such gases include $CH_3Cl$ (methyl chloride), $CH_3Br$ (methyl bromide), $CH_3I$ (methyl iodide), $CH_3OH$ (methanol), and terpenes. Methyl chloride, in particular, has been shown in a previous study to have a longer photochemical lifetime in the atmospheres of Earth-like planets orbiting M dwarfs compared to sunlike stars (Segura et al. 2005), though note that the other methyl-containing compounds listed here were not tested in the Segura et al. study. Natural sources of methyl chloride, the most abundant halocarbon in the atmosphere (Yokouchi et al. 2000), include oceanic sources (Koppmann et al. 1993) such as planktonic algae (Harper et al. 2003), plants (Rhew 2011; Yokouchi et al. 2000), and biomass burning (Blake et al. 1996). Higher order halogenated organic compounds such as $C_2H_4Cl$, $C_3H_7Cl$, and $C_4H_9Cl$, as well as methyl chloride itself, can also be produced in soils and sediments when chlorine ions are alkylated through organic matter oxidation (Keppler et al. 2000). An overview of methyl chloride sources, including industrial ones, is provided in Keppler et al. (2005). Non-anthropogenic sources of methyl bromide include marine algae and biomass burning (McCauley et al. 1999; Blake et al. 1996), kelp (Manley & Dastoor 1988), other oceanic sources (Anbar et al. 1996), and plants (Rhew 2011).

Methyl iodide can be produced by vegetation and soils (Sive et al. 2007), oceanic sources (Rasmussen et al. 1982) including kelp and microbial metabolisms (Manley & Dastoor 1988), and biomass burning (Blake et al. 1996). Methanol is the simplest alcohol and is produced by a variety of anaerobic metabolisms including methanotrophy (Xin et al. 2004). In the atmosphere, it can contribute to the tropospheric HOx budget after being oxidized to formaldehyde (Solomon et al. 2005). Terpenes are a broad class of organic compounds released by plants and insects (Pare & Tumlinson 1999) that are responsible for atmospheric phenomena such as the low-lying blue haze that can be seen over some forested regions like the Blue Ridge Mountains (Went 1960). Some terpenes such as d-limonene can ultimately generate gases like formaldehyde and formic acid in the atmosphere through photochemical processing (Walser et al. 2007).

The largest potential fluxes of these gases would likely be on planets where they are the byproduct of metabolism. On modern Earth, major sources of these gases (including $CH_3SH$) are typically the degradation of amino acids (Stoner et al. 1994; Lomans et al. 2002). However, on planets with a different atmospheric composition, there might be an energetic incentive to produce these gases directly. The potential for this has been demonstrated in the laboratory, in experiments where methanogens are given $H_2S$ as a substrate in place of $H_2$, and in response produce $CH_3SH$ instead of $CH_4$ (Moran et al. 2008). On a planet with globally-high $H_2S$ concentrations, microbes could release high fluxes of $CH_3SH$ to the atmosphere, accumulating $CH_3SH$ and its photochemical byproducts ($C_2H_6$ in particular) to potentially detectable levels (Domagal-Goldman et al. 2011). Other atmospheric compositions may also lead to an incentive for other gases to be produced. For example, an atmosphere rich in halogens - and HCl, HI, and HBr in particular - may incentivize the biological production of $CH_3Cl$, $CH_3I$, and $CH_3Br$. While this has not been studied in the laboratory, nor modeled in a photochemical simulation, this sort of atmosphere would create the largest concentration of - and therefore spectral signal from - these gases.

Such gases have been considered in the context of atmospheric biosignatures since they are produced by biological processes (Seager et al. 2012), although as we have seen here and in Domagal-Goldman et al. (2011), photochemistry and the strength of these gases' spectral signatures will determine whether they – or any of their photochemical byproducts – are useful gases (or aerosols) to target in the search for life beyond Earth. For instance, the spectral signatures of $CH_3Cl$ were investigated in Segura et al. (2005) in simulations that included photochemistry. $CH_3Cl$ was readily destroyed by the solar spectrum for a planet orbiting the sun, but for planets orbiting M dwarfs AD Leo or GJ 463C, the $CH_3Cl$ vertical mixing ratio profiles approximated isoprofiles up to at least 70 km in altitude. For such atmospheres, Segura et al. (2005) showed that $CH_3Cl$ produces absorption features in the thermal infrared near 7, 10, and 15 $\mu$m. Although Segura et al. (2005) did not simulate transit transmission spectra, the well-mixed profiles of $CH_3Cl$ for the M dwarfs suggest it may be detectable in transit observations, even if such observations could only probe the upper layers of these atmospheres.

4.3 Interpretation of Biosignatures in the Planetary Context

Generally speaking, it is important to consider any potential biosignature in the broader planetary context to rule out false positive abiotic production mechanisms and search for other signs of habitability and life. Such additional constraints are important in all situations: the best understanding of any planet will come from a holistic consideration of many diverse pieces of information. The solar system shows us that there is tremendous value in obtaining as much information as possible when interpreting difficult-to-access planetary characteristics. For example, the case for Europa's subsurface ocean was supported and strengthened by multiple lines of evidence ranging from surface morphology to Europa's induced magnetic field (e.g. Squyres et al. 1983; Carr et al. 1998; Kivelson et al. 2000; Stevenson 2000).

Alarmingly, the nearby worlds in our solar system also have a rich history of misinterpretation in the absence of robust data. Percival Lowell's extensive analysis of the putative (and wholly illusionary) canals on Mars is perhaps the best-known example of this phenomenon (Lowell 1906), but there are countless other less notorious cases. For instance, when G. Kuiper discovered Titan's atmosphere (Kuiper 1944), he speculated that the moon's orange coloration is due to "the action of the atmosphere on the surface itself, analogous to the oxidation supposed to be responsible for the orange color of Mars," not understanding that he was seeing the atmosphere rather than the planetary surface. Exoplanets, spatially unresolved and at vast distances, will be even more difficult to interpret. Because the designs of the first generation of observatories that include terrestrial exoplanet characterization and biosignature detection in their science goals are currently being studied, it is crucial that these design studies be informed by ongoing analyses of a variety of possible biosignatures, habitability signatures, and false positives of both. For these future observatories, a wide wavelength spectral range can mitigate the possibility of reaching erroneous conclusions by providing additional spectral context to consider any given feature. Additional context will be even more valuable when interpreting newly discovered planet types and characteristics not represented in the solar system (e.g. hot Jupiters, mini-Neptunes, and super Earths). The diversity of planet types already discovered suggest that the probability of detecting a true Earth twin is very small, so it is crucial to expand our understanding of how habitability and biosignatures may appear on worlds different from the planet we life on. Somewhat paradoxically, we must do this before these exoplanets are even discovered; otherwise, we risk under-designing the capabilities of the future observatories that will study them.

Organic haze is just one example of a potential novel biosignature. Haze itself is relatively simple to detect due to its strong spectral features, but as we have argued here, interpretation of haze spectral features will be challenging, dependent on several different pieces of information. Indeed, Titan's abiotic organic haze shows plainly that the detection of haze alone is inadequate for biosignature considerations. We have, in this study, discussed observing strategies to strengthen the case for a haze as biogenic: 1) When a haze is present in a sufficiently oxidizing background atmosphere, high methane fluxes may be required to produce it, and such high fluxes may suggest (but not prove) the involvement of biology, the most vigorous producer of methane on Earth today. 2) When a haze is present in an atmosphere that has less methane than photochemical models predict is required to initiate haze formation, this implies the existence of an additional hydrocarbon source like $S_{org}$ gases. Beyond these two strategies, other measurements like the presence (or absence) of other biosignatures and habitability signatures like ocean glint (Robinson et al. 2010; Robinson et al. 2014) would provide valuable additional information that could strengthen (or weaken) the case for biological involvement.

Future work is necessary to more rigorously examine the rates of abiotic methane production that may be possible on other planet types and determine whether there are other abiotic mechanisms that generate hazes at unexpectedly low $CH_4/CO_2$ ratios. Additionally, as we emphasized above, future studies of other novel biosignatures and habitability signatures are also needed to expand our palette of known possible types of living planets. However, because photochemistry may generate non-intuitive or indirect spectral signs of these biological processes (e.g. the spectral impacts of $S_{org}$), it is crucial for any considerations of novel spectral signatures to also include the context of the broader atmospheric and stellar flux environment.

4.4 False Positives for Organic Haze

We considered several possible spectral mimics for organic haze in our previous study (Arney et al. 2017). Given that the most detectable spectral signature of organic haze is its blue-UV absorption feature, other compounds with strong UV-blue absorption have the potential to be mistaken for organic haze. These include iron oxide, the unknown UV absorber in the Venusian atmosphere, and exotic haze compounds such as zinc sulfide (ZnS, although its high condensation temperature means it is not a viable aerosol candidate for Earth-like atmospheres). An additional UV-blue-absorbing compound we did not consider in our previous study is $S_8$ particles, which can produce deep, broad blue-UV absorption features similar to organic haze when present in large quantities (Hu et al. 2013). $S_8$ particles are produced by volcanic sulfur gas emissions; photochemical processing of these volcanic gases tends to favor production of $H_2SO_4$ aerosols under more oxidizing conditions, while $S_8$ is favored under more reducing conditions such as those that would also tend to generate organic haze (e.g. Hu et al 2013, Zahnle et al 2006). It may be possible to distinguish the spectral signatures of $S_8$ from organic haze via absorption features from emitted volcanic gases themselves (e.g. $H_2S$ and $SO_2$), but modeling work by Hu et al. (2013) shows these gases are photochemically short-lived, and their most detectable absorption features occur longward of 5 µm. This makes direct imaging detections of them very difficult – although transit observations probing to longer wavelengths may still be able to measure them. Organic haze produces its own diagnostic absorption features near 3 and 6 µm that may also be detectable in transit observations, allowing it to be distinguished from $S_8$ if infrared transit observations are possible. We emphasize, again, as we did in Section 4.3, that any given spectral feature must be considered in the context of the whole planetary environment and with as much spectral contextual information as possible. A strong UV-blue wavelength absorber is more likely to be organic haze than $S_8$ particles if it is detected in the presence of strong $CH_4$ features, but if $CH_4$ features are weak or absent, $S_8$ (or other compounds like iron oxide) may be more likely. $S_8$ (and sulfate) concentrations may also be time-variable if volcanic outbursts occur sporadically, and time-resolved spectroscopy may be a powerful means of identifying false positives of volcanic emissions (Misra et al. 2015).

5. Conclusions

Organic haze formation on Archean Earth was likely controlled by biological methane production, and this type of haze may also occur on anoxic worlds elsewhere. On planets with Archean-like $CO_2$ levels, organic haze formation requires methane production consistent with known and theoretical biological fluxes on Earth. However, because abiotic processes can also produce methane, methods to distinguish biotic and abiotoic hazes are needed. To that end, we explored how biogenic organic sulfur gases affect haze formation since these gases can liberate methyl radicals that become involved in haze production. We find organic sulfur gases can drive haze formation at lower $CH_4/CO_2$ ratios compared to methane photochemistry alone. This effect is especially pronounced around M dwarfs with lower UV fluxes than the sun. To make the case for a $S_{org}$-mediated haze impacting a planet's spectrum, it will be necessary to constrain the atmospheric $CH_4/CO_2$ ratio to test whether the haze is present at an anomalously low ratio unexplainable by methane photochemistry alone. Although $S_{org}$ gases themselves are difficult to detect in a planet's spectrum, organic haze produces a very strong absorption feature at UV-blue wavelengths. This haze could also be detected in the NIR and mid-IR using transit transmission observations. Methane and carbon dioxide produce absorption features in the NIR that could be detected in both reflected light and transit transmission observations. Future observatories and telescope concepts such as JWST, LUVOIR, and HabEx may be able to make measurements of these spectral features. Because the haze absorption feature is so strong, it may be one of the most detectable spectral beacons of life, although interpreting the haze absorption feature in the context of its environment may be more challenging. The long anoxic history of our planet teaches us not to ignore the challenge of life detection on anoxic worlds when we design the instruments, plan the observations, and search for life on distant exoplanets.

Tables

| Mirror Diameter (m) | cutoff for IWA = λ/D (μm) | | cutoff for IWA = 2 λ/D (μm) | | cutoff for IWA = 3 λ/D (μm) | |
|---|---|---|---|---|---|---|
| | GJ 876 | Sun | GJ 876 | Sun | GJ 876 | Sun |
| 12.7 | 1.6 | 14 | 0.82 | 6.8 | 0.54 | 4.6 |
| 7.6 | 0.98 | 8.2 | 0.49 | 4.1 | 0.33 | 2.7 |
| 5.5 | 0.71 | 6.0 | 0.36 | 3.0 | 0.24 | 2.0 |
| 4 | 0.52 | 4.3 | 0.26 | 2.2 | 0.17 | 1.4 |

Table 1. Wavelength cutoffs for four mirror diameters and three IWAs for planet-star systems as 4.5 pc.

Figures

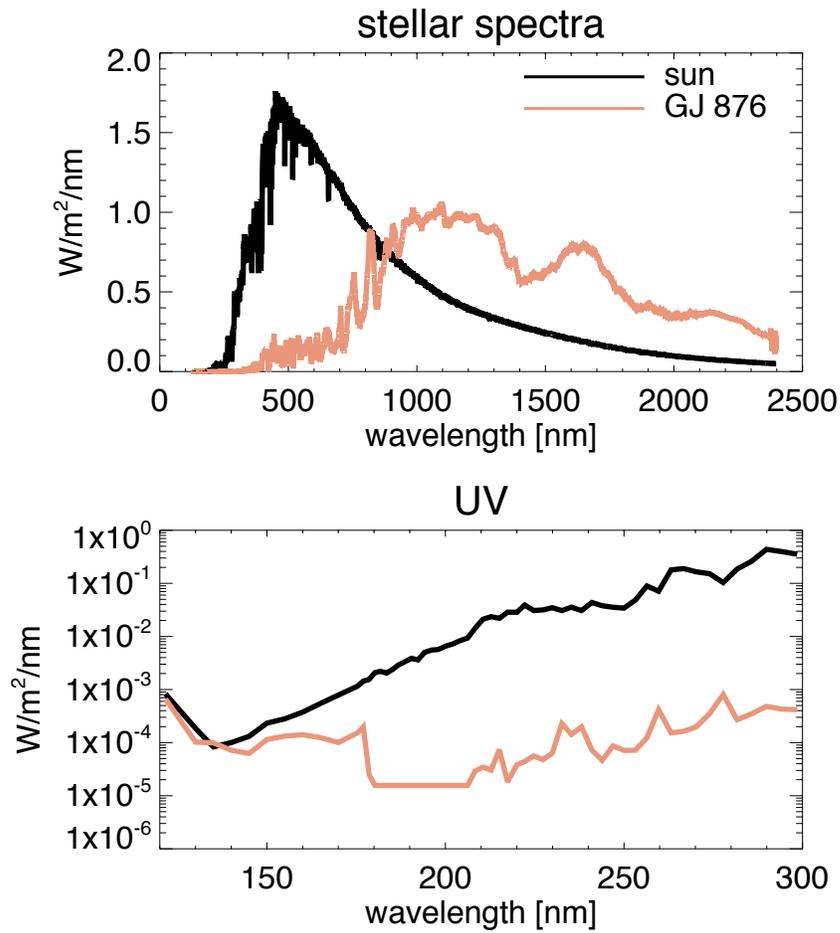

Figure 1

Spectra of the Archean sun and GJ 876 used in this study. The bottom panel shows the actual UV wavelength grid used in the photochemical model. Wavelength bins are larger than individual emission lines (e.g. Lyman α).

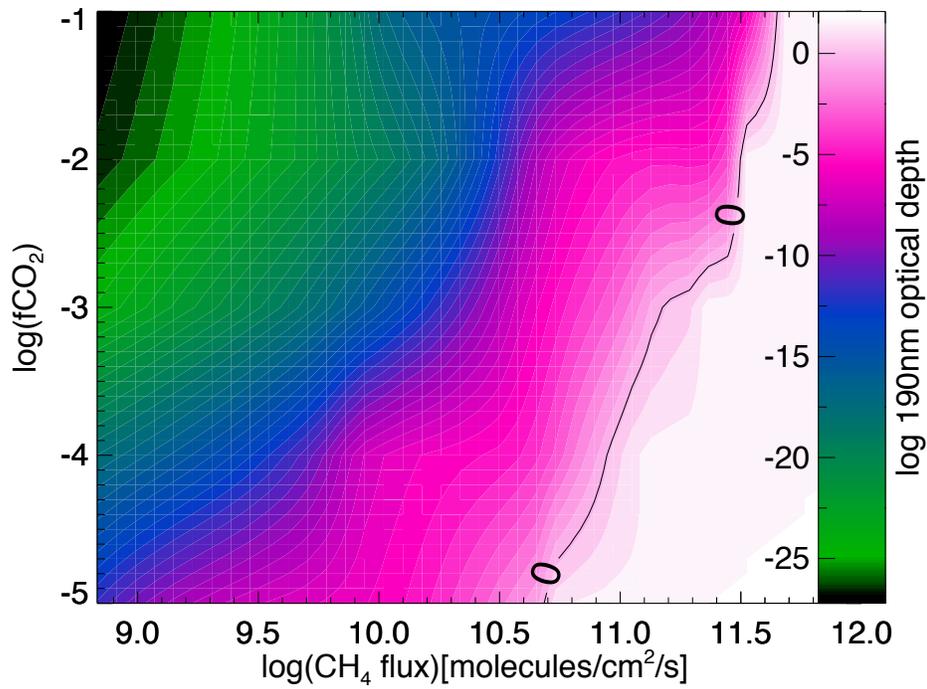

Figure 2

The colored contours show the log optical depth of the Archean haze (for a planet orbiting the sun) at 190 nm as a function of the log atmospheric fraction of $CO_2$, $\log(fCO_2)$, on the vertical axis, and the log of the $CH_4$ surface flux on the horizontal axis. Haze optical depth of unity ($\log(1) = 0$) is marked by the solid black line.

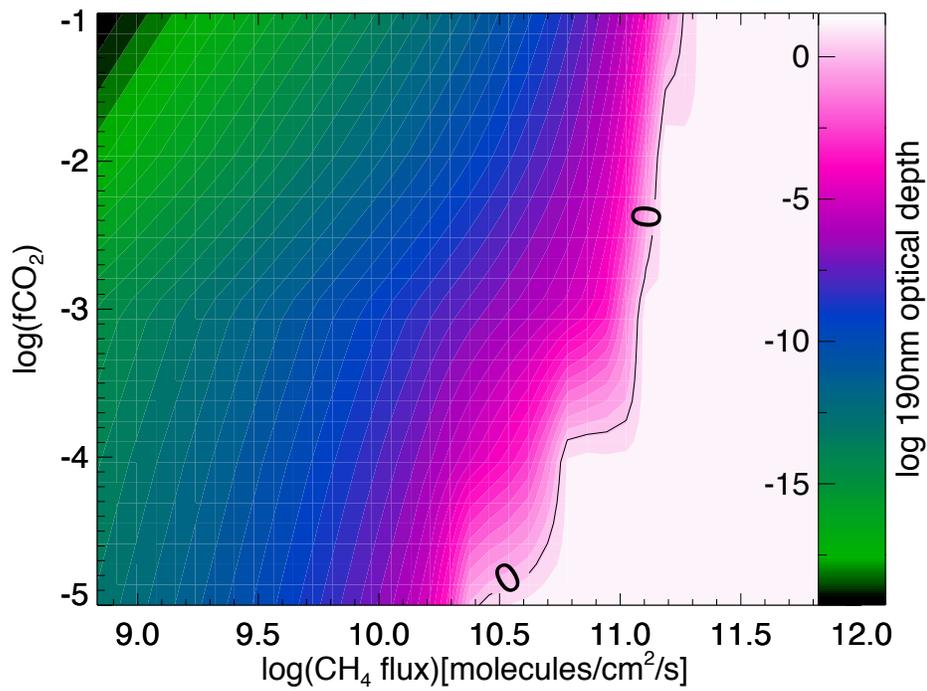

Figure 3

Same as Figure 3, but for an Archean-analog planet orbiting GJ 876.

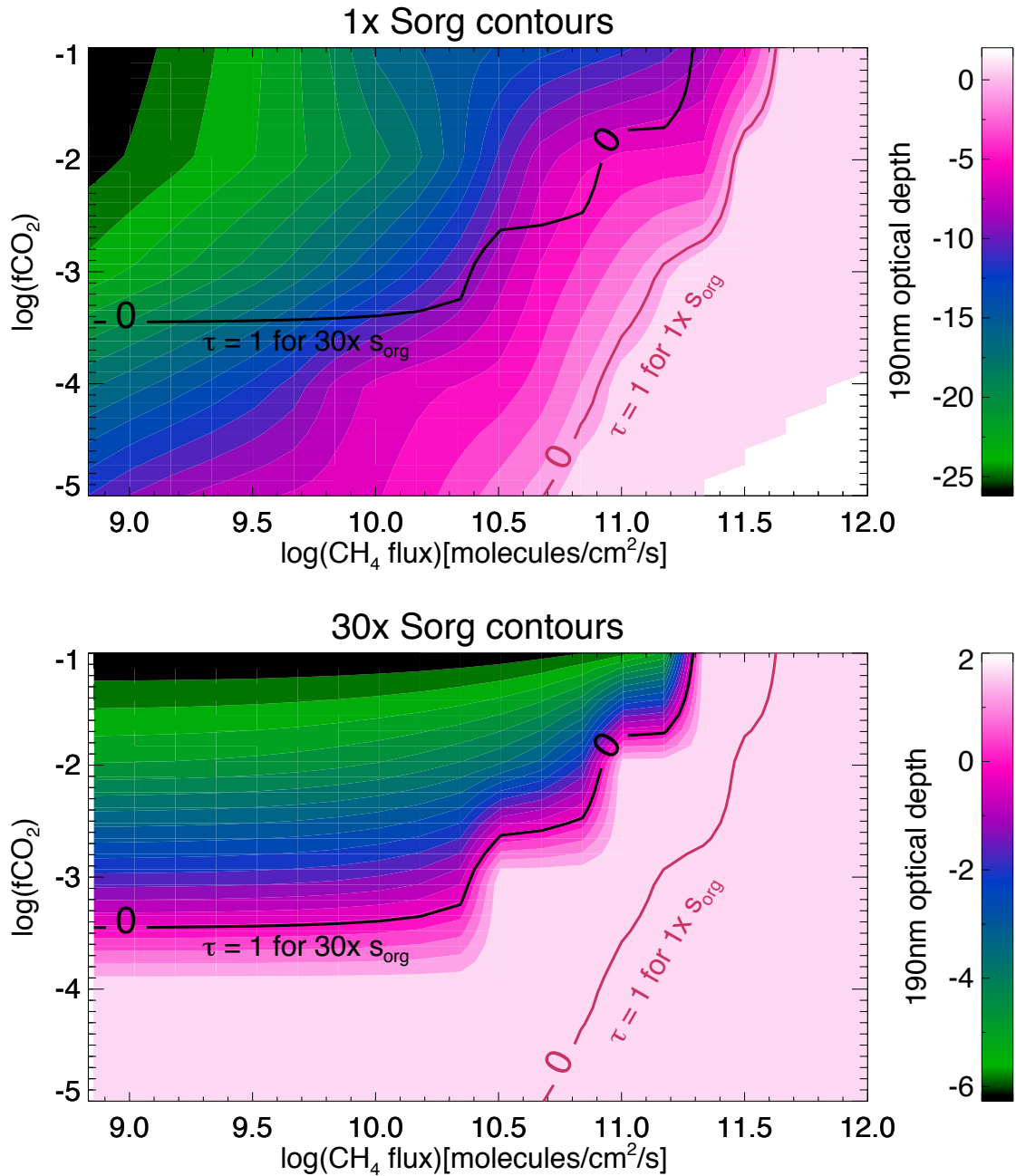

Figure 4

The log of the 190 nm optical depth of organic haze for planets around the sun at 1 x $S_{org}$ and 30 x $S_{org}$ as a function of log($fCO_2$) and surface $CH_4$ flux The red line in both panels shows where optical depth is unity for 1 x $S_{org}$ and the black line in both panels shows where optical depth is unity for 30 x $S_{org}$ The 1x and 30x $S_{org}$ optical depth of unity lines are overlain over the contours on both panels so that they may be easily compared.

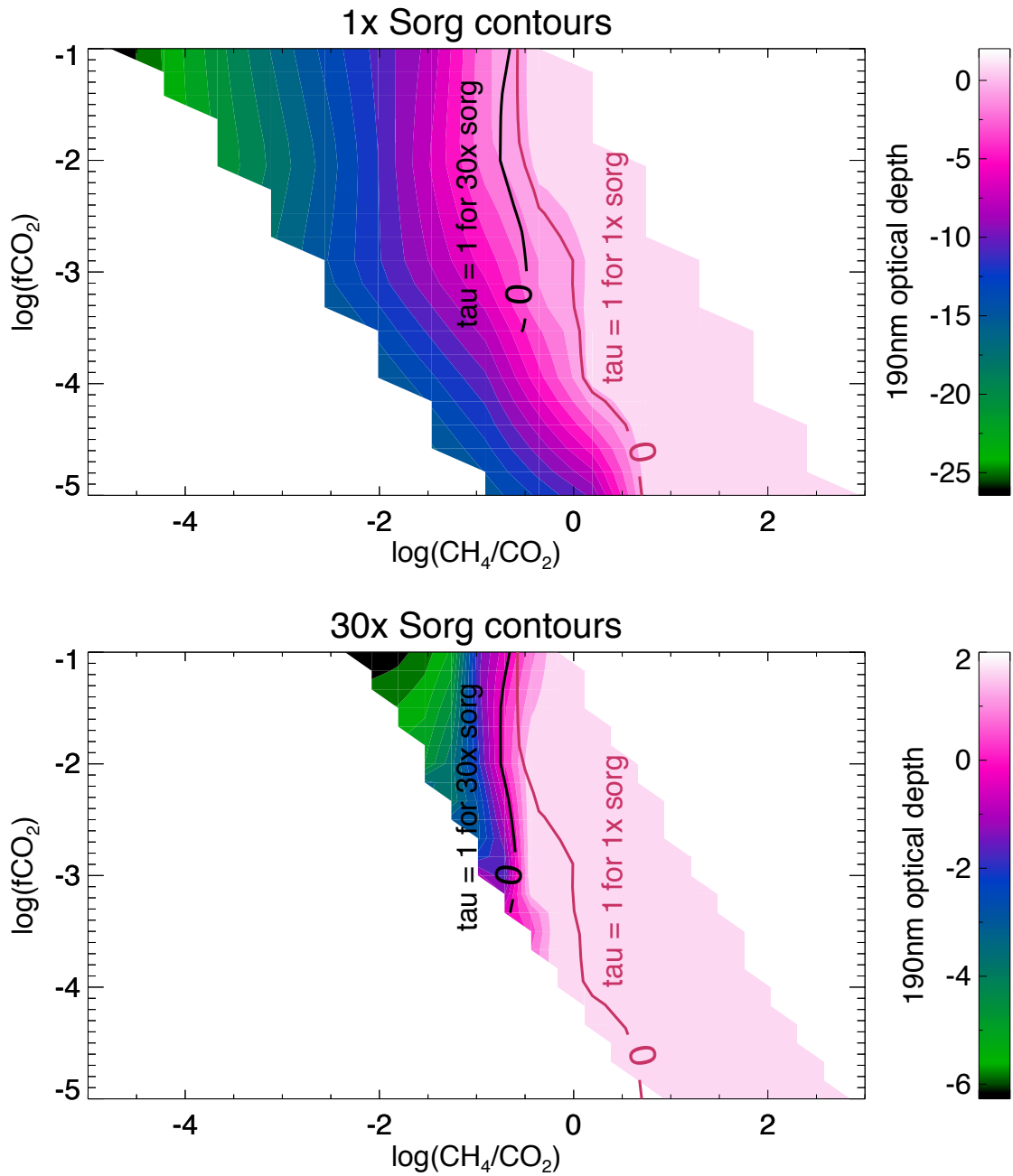

Figure 5

Same as Figure 4, but showing the log of the 190 nm optical depth of organic haze for planets around the sun at 1 x $S_{org}$ and 30 x $S_{org}$ as a function of $\log(fCO_2)$ and $\log(CH_4/CO_2)$.

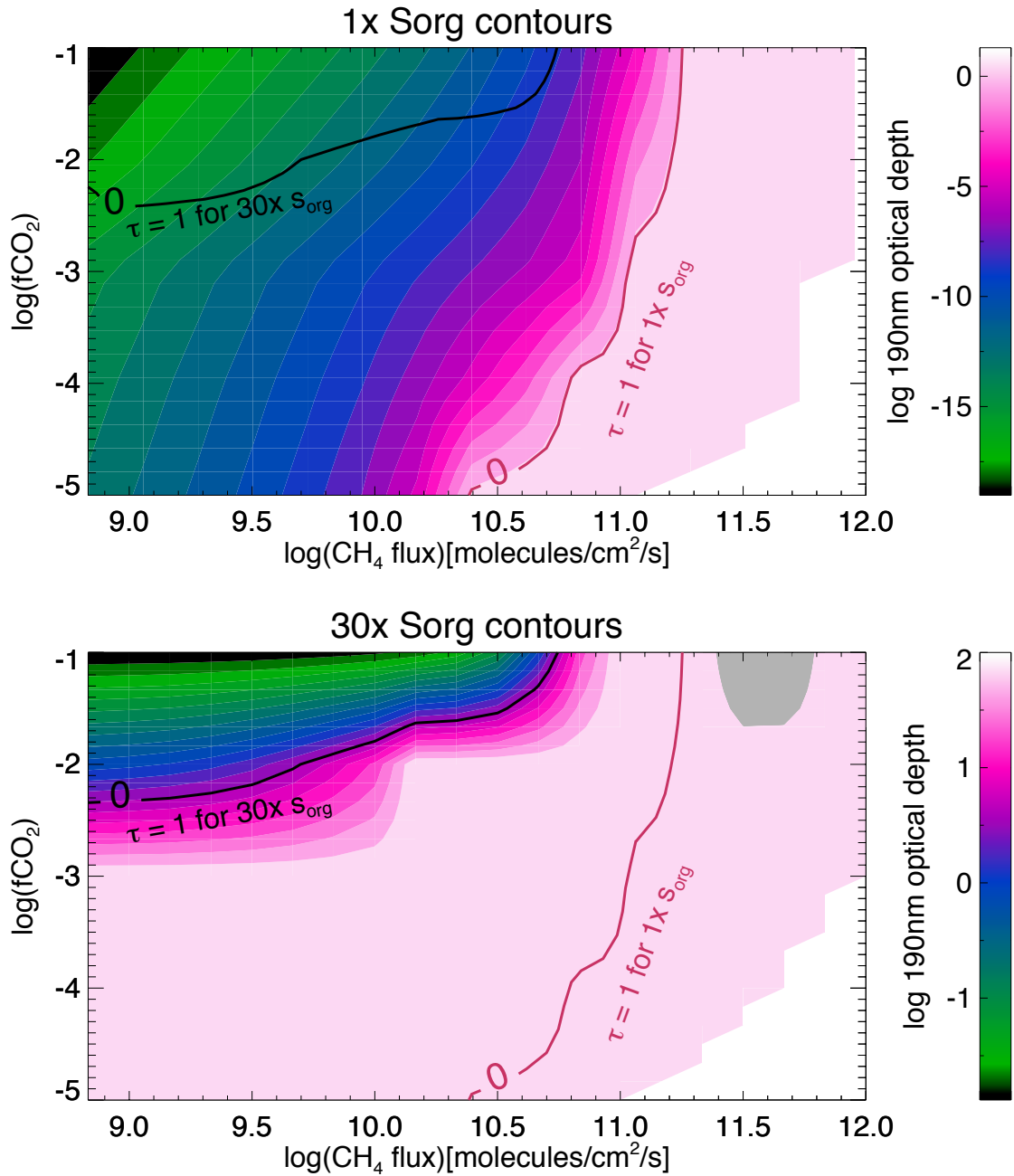

Figure 6

Same as Figure 4, but for planets orbiting GJ 876. Simulations in the gray region are not converged.

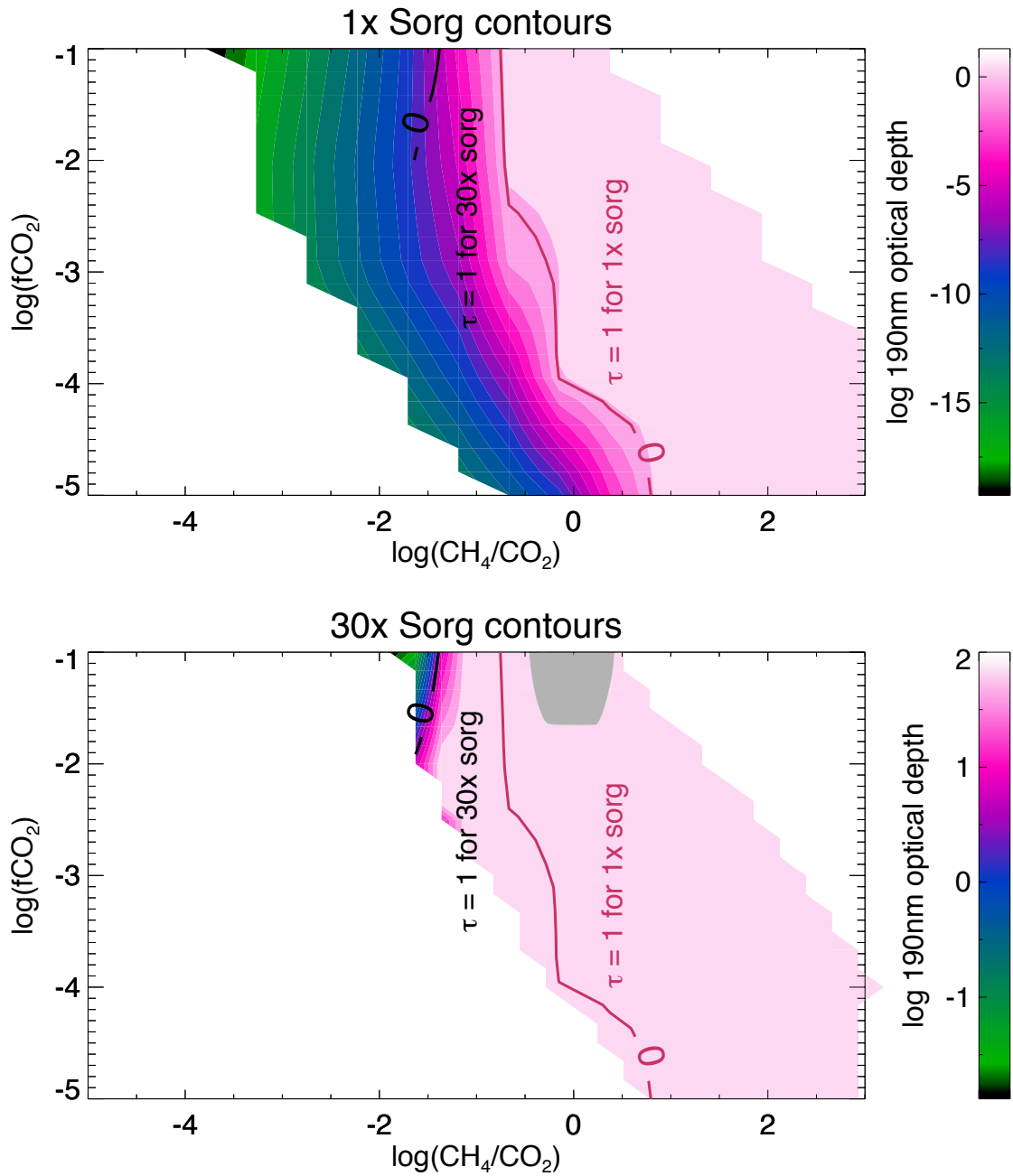

Figure 7

Same as Figure 5, but for planets orbiting GJ 876.

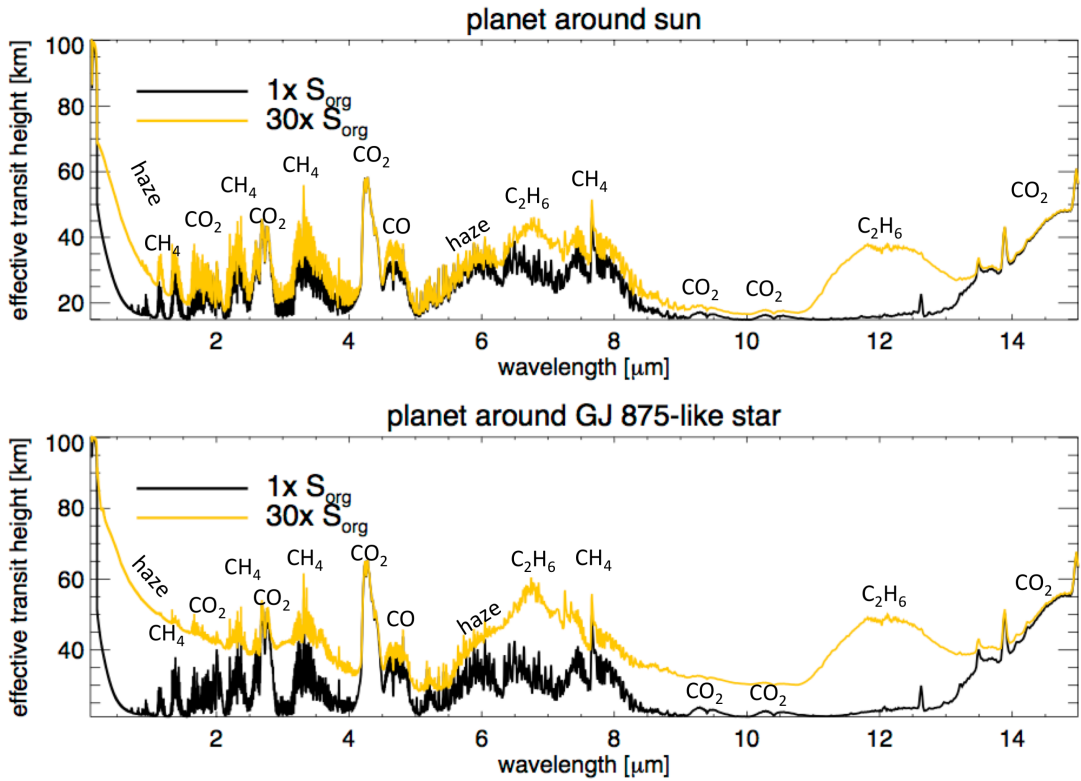

Figure 8

Representative transit transmission spectra of Archean Earth-like planes with different $S_{org}$ fluxes. The y-axis shows the effective tangent height, which is the altitude above the planet's surface that light on tangent transit path lengths can penetrate into the atmosphere. Spectra are shown for $\Delta\lambda = 0.01$ μm.

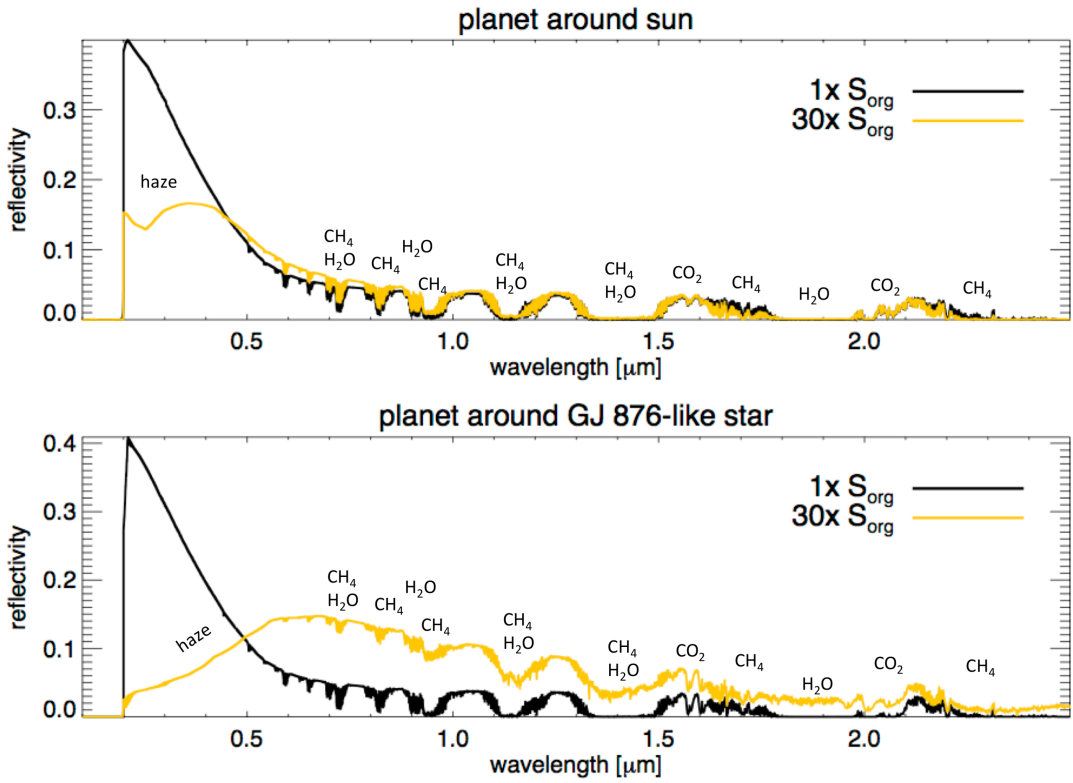

Figure 9

Representative reflected light spectra of Archean Earth-like planes with different S$_{org}$ amounts for the same atmospheres shown in Figure 8. These spectra do not include water clouds to show the spectral impact of only organic haze.

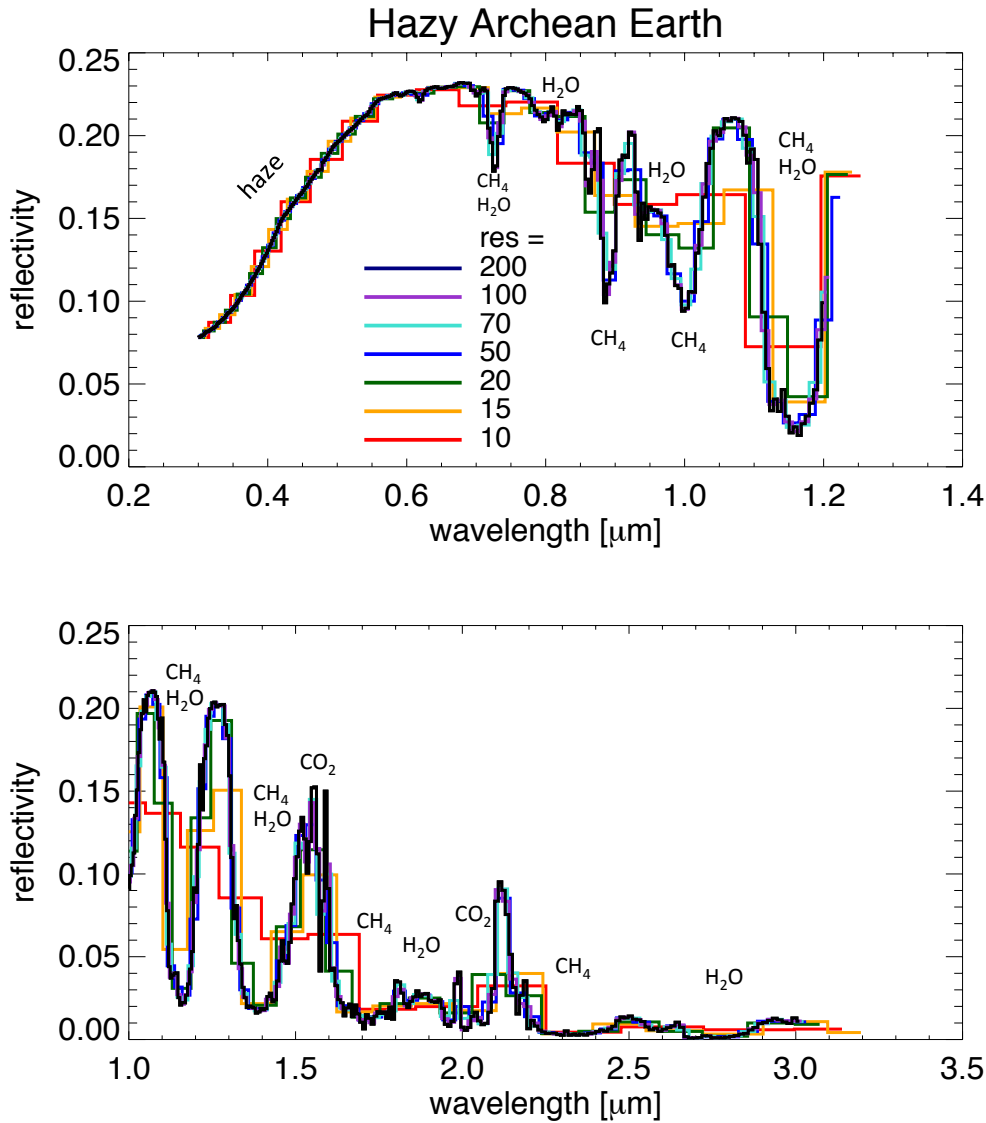

Figure 10

The reflected light spectra of hazy Archean Earth at several spectroscopic resolutions. The haze absorption feature at λ < 0.6 μm is sufficiently strong and broad to be resolved at very low spectral resolution. These spectra include water clouds in addition to haze, via a weighted averaging technique in our 1D model (50% haze only, 25% haze and cirrus cloud, and 25% haze and stratocumulus cloud).

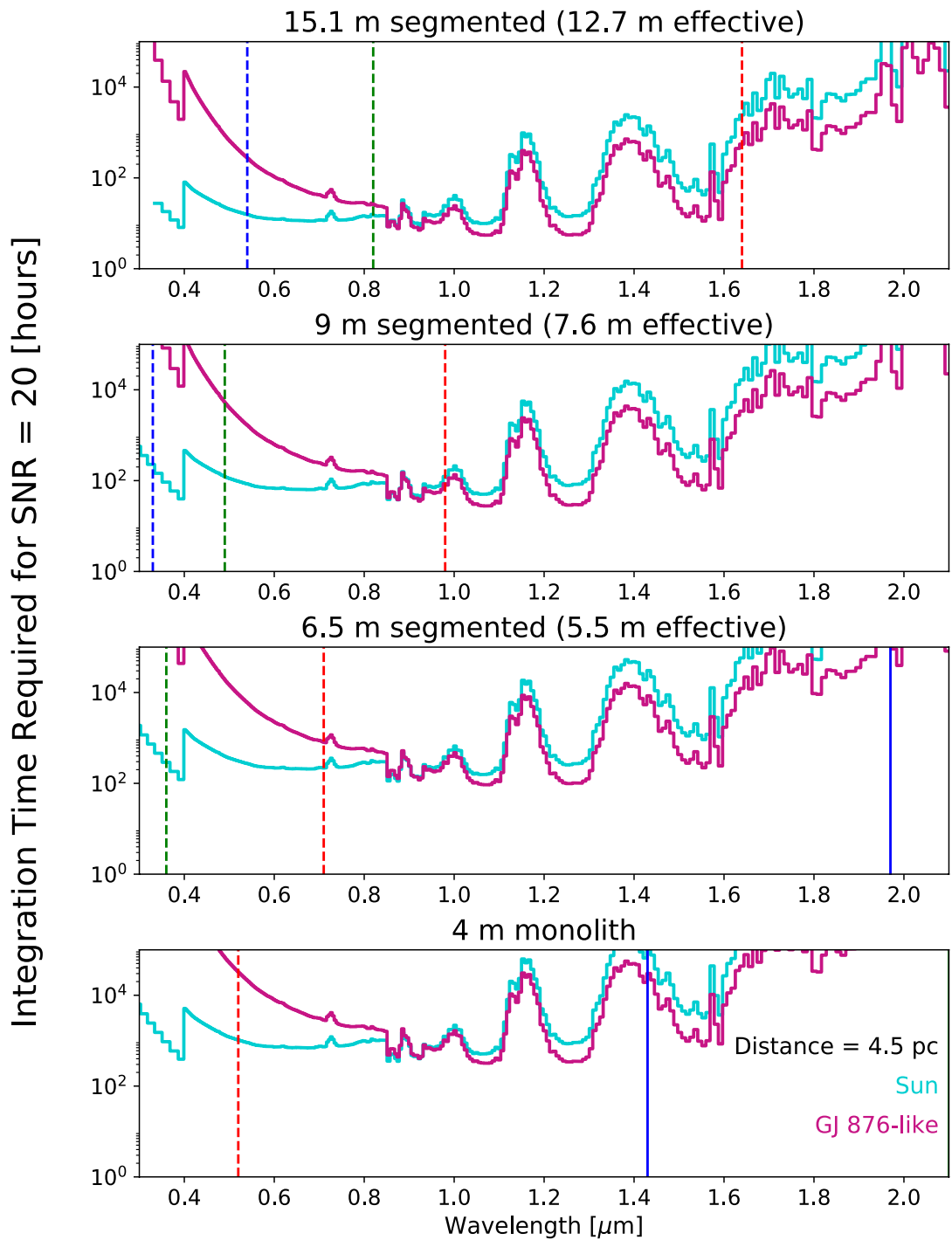

Figure 11

The wavelength-dependent integration time required to obtain a signal-to-noise ratio (SNR) = 20 for hazy Archean-analog planets orbiting the Sun and GJ 876 at 4.5 pc as observed by four different telescope architectures. The red, green, and blue vertical lines represent the wavelength cutoffs for IWA = $\lambda/D$, 2 $\lambda/D$, and 3 $\lambda/D$, respectively. The dashed lines are the IWA cutoffs for a GJ 876-like star, and the solid lines represent the IWA cutoffs for a solar-type star. Not all IWA cutoffs are shown on every plot if they do not overlap with the wavelength range displayed: for instance, all of the solar IWA lines occur redward of the right axis of the 12.7 and 7.6 m plots. The discontinuity at 0.4 μm and 0.85 μm is due to the boundary between the assumed UV, visible, and NIR detectors.

Acknowledgments

This work was performed as part of the NASA Astrobiology Institute's Virtual Planetary Laboratory, supported by the National Aeronautics and Space Administration through the NASA Astrobiology Institute under solicitation NNH12ZDA002C and Cooperative Agreement Number NNA13AA93A. Simulations were facilitated through the use of the Hyak supercomputer system at the University of Washington eScience Institute. We thank Dr. T. McCollom for pointing us to sources discussing methane in serpentinizing systems. Spectra shown in this work will be archived at the Virtual Planetary Laboratory online spectral database. We thank our two anonymous reviewers for their helpful comments and suggestions that improved our manuscript.

Author Disclosure Statement

No competing financial interests exist.